\documentclass[12pt]{article}

\usepackage{scicite}
\usepackage{times}
\usepackage{graphicx}
\usepackage{caption}
\newenvironment{sciabstract}{%
\begin{quote} \bf}
{\end{quote}}

\renewcommand{\figurename}{{Fig~}}

\title{Topological transport of a classical droplet in a lattice of time}

\author
{Tapio Simula,$^{1}$ Niels Kj{\ae}rgaard,$^{2}$ Tilman Pfau$^{3}$\\
\\
\normalsize{$^{1}$Optical Sciences Centre, Swinburne University of Technology, Melbourne 3122, Australia}\\
\normalsize{$^{2}$Department of Physics, QSO---Centre for Quantum Science,}\\
\normalsize{and Dodd-Walls Centre for Photonic and Quantum Technologies,}\\
\normalsize{University of Otago, Dunedin, New Zealand}\\
\normalsize{$^{3}$5. Physikalisches Institut, Universit\"at Stuttgart, Pfaffenwaldring 57, 70569 Stuttgart, Germany}\\
\\
}

\date{}

\begin{document} 


\baselineskip24pt


\maketitle


\begin{sciabstract}


Thouless charge pumps are quantum mechanical devices whose operation relies on topology. They provide the means for transporting quantum matter in space lattices with a single quantum precision. Contrasting space crystals that spontaneously break a continuous spatial translation symmetry and form crystals in space,
time crystals have emerged as novel states of matter that organize into time lattices and spontaneously break a discrete time translation symmetry. 
The utility of Thouless pumps that enable topologically protected quantised transport of electrons and neutral atoms in spatial superlattices leads to the question if corresponding devices exist for time crystals?
Here we show that topological pumps can be realized for time solids by transporting droplets of a liquid forward and backward in time lattices and we measure the topological index that characterises such pumping processes. 
By exploiting a synthetic time dimension classical time crystals can circumvent the quantum tunneling that underpins Thouless charge pumps.
Our results establish topological pumping through time instead of space and pave the way for applications of time crystals.
\end{sciabstract}


\section*{Introduction}
Conservation laws of momentum and energy are deeply rooted in the symmetries of space and time translations \cite{Noether1918a}. The spontaneous formation of spatial crystals, such as perfect diamonds, break \emph{continuous} space translation symmetry and as a consequence linear momentum of a particle traveling through the lattice is not conserved. While similar crystals in time \cite{Wilczek2012a,Shapere2012a} may not exist, \emph{discrete} time crystals that emerge in periodically driven systems \cite{Sacha2018a,Zaletel2023a} have been realized in recent ground breaking experiments \cite{Zhang2017a,Choi2017a,Rovny2018a}. Within this new paradigm, the spontaneous breaking of the discrete time translation symmetry (TTS) is manifested as the system's sub-harmonic response to the driving \cite{Sacha2015a}. In the realm of quantum mechanics discrete time crystals emerge as non-equilibrium many-body phases of matter that have recently been studied using NISQ era quantum computers \cite{google,Frey2022a}. Theoretical works \cite{Hannaford2022a} have also considered solid state physics time-analogues ranging from time localisation \cite{Sacha2015a} and time crystal insulators \cite{Giergiel2019a} to time-space pumps \cite{Braver2022a}. As has been pointed out \cite{Yao2020a}, sub-harmonic responses are also found in a range of periodically driven classical physical systems. Many of such classical systems, however, lack inherent stability and a small amount of noise will destroy their period doubled oscillations. The salient and crucial feature of the classical system studied here is that it does not suffer from such instabilities.

\begin{figure*}[t!]
\centering
\includegraphics[width=\textwidth]{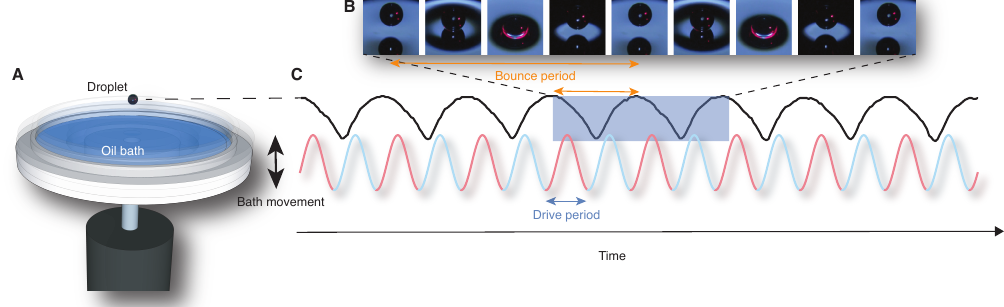}
\caption{{\bf Bouncing droplet time crystal.} (\textbf{A}), Oil droplet deposited on a bath driven by a vertical sinusoidal drive. (\textbf{B}), Images of droplet shown at 6.25~ms intervals revealing repetition every 4th frame as a result of the system's underlying 25~ms discrete TTS, contrasting with the 12.5m~s TTS of the drive. (\textbf{C}), Vertical droplet position $h(t)$ (black line) inferred from the images acquired using high-speed camera. The period of the droplet's bouncing motion is exactly twice that of the applied sinusoidal drive of the bath shown in dual red/blue color to emphasize the broken TTS of the droplet motion. The lattice parameters are $V_s=2.93 g$ and $V_l=0$.}
\label{fg:setup}
\end{figure*}

\section*{Creation of time crystals}

Figure \ref{fg:setup}A shows our experimental setup, where an oil droplet bounces on the surface of an oil bath driven in an up and down motion (see Movie, and Materials and Methods). Remarkably, when a droplet of a viscous fluid is deposited on a driven bath of the same fluid, its coalescence may be inhibited allowing the droplet to continue bouncing indefinitely \cite{Couder2005,Bush2021a,Saenz2021a}. And, within a broad window of drive amplitudes and frequencies the droplet may spontaneously undergo a discrete TTS breaking and will begin to skip every second period of the drive. In our system \cite{Simula2023a}
such period-doubled bouncing is observed when the bath is shaken sinusoidally along the axis of gravity at a fundamental driving frequency $f=80\;$Hz with a corresponding period $T=1/f=12.5$~ms. Figure~\ref{fg:setup}B shows a stroboscopic sequence of photos of such a droplet sampled at frequency $2f$, revealing every fourth image to be near indistinguishable and the droplet bouncing to have a period $2T=25$~ms. The emergent $2T$ periodicity and the associated so-called crypto-equilibrium \cite{Zaletel2023a} are defining characteristics of discrete time-crystals. Such period doubled bouncing took place already in the earliest bouncing droplet experiments \cite{Couder2005} but has only recently become associated with time crystal behaviour \cite{Simula2023a}. Figure~\ref{fg:setup}C presents the vertical height $h(t)$ of the droplet (black) together with the acceleration of the bath as functions of time $t$. The height is tracked by illuminating the droplet with a red laser beam and detecting the reflection. It is evident that one of the two equivalent bouncing modes (in this case A (blue) parity mode) is singled out by the spontaneously broken TTS. We have verified that the period doubled bouncing is robust with respect to perturbations and is stable over extremely long time scales, and we attribute this quality to the absence of `time phonons' (see Materials and Methods, and Supplementary Figure~S1).

\begin{figure*}[t!]
\centering
\includegraphics[width=\textwidth]{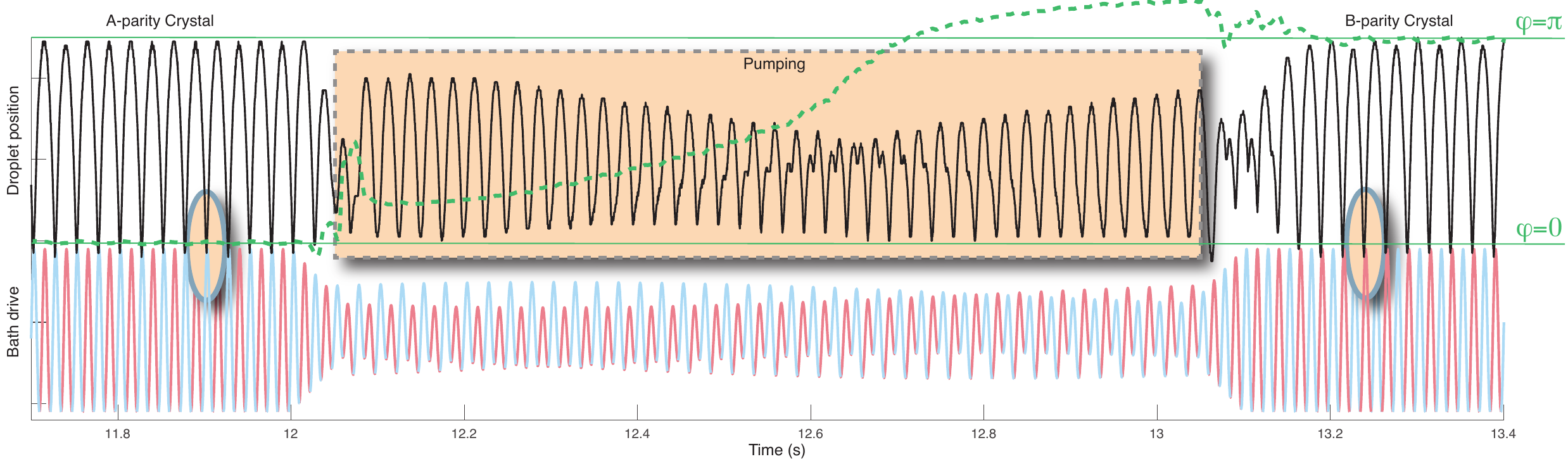}
\caption{{\bf Parity conversion.} Droplet position $h(t)$ as a function of time (solid black) and the acquired time-geometric phase (dashed green) with the A-parity, B-parity, and pump-on interval highlighted (ochre). The driving waveform $a(t)$ with blue and red colours as in Fig.1. The details of the drive parameters are provided in Methods.}
\label{fg:parity}
\end{figure*}

\section*{Time crystal conversion}
We can readily create and observe droplet time crystals on demand, and in any given realization once the system settles into a type A or type B time crystal, it remains locked to its type. In space crystals, Thouless pumps provide deterministic and topologically protected site-to-site transport for a quantum system \cite{Thouless1983a,Niu1984a,Lohse2016a,Nakajima2016a,Lu2016a}. This raises the question if a similar topological pumping mechanism exists for transporting our classical droplet time crystal. To achieve this, we first prepare a droplet time crystal illustrated in Fig.~\ref{fg:setup} by setting the driven bath acceleration to be a pure sinusoid $a(t) = a_{80}(t)$ where $a_{80}(t)= g + V_s\sin(2\pi f t)$ with $g$ the constant acceleration due to gravity and $V_s$ the amplitude of a `short-time lattice'. We convert between the two time crystal types by superimposing an additional long-time lattice---a sinusoid of twice the period of the short-time lattice---to yield a superlattice bath acceleration $a(t) = a_{80}(t)+a_{40}(t)$, where $a_{40}(t)= V_l\sin(\pi f t +\theta(t))$. Here $V_l$ is the amplitude of the long time lattice, and $\theta(t)$ is a phase that we sweep linearly in time. 

The blue/red curve in Fig.~\ref{fg:parity} shows the full synthesised bath acceleration $a(t)$ time-series that realizes the desired time crystal type conversion (Methods). Initially, $V_s = 2.93 g$ and the droplet (black curve) is in a time crystal phase undergoing period doubled bouncing in the A (blue) parity mode. Turning on the long lattice by smoothly changing the parameters to $V_s = 1.28 g$ and $V_l = 0.46 g$, followed by a one-second linear ramp of $\theta$ from zero to $\pi$, induces the droplet to undergo a parity swap type conversion such that at the end of the phase ramp it is observed to be bouncing in the B (red) parity mode. The ochre shading in the middle of Fig.~\ref{fg:parity} highlights the droplet's bouncing in a superposition of the A and B parities. To quantify the dynamics of the deterministically induced A-to-B conversion we performed a 40~Hz quadrature (Q) and in-phase (I) demodulation (Methods) of the measured droplet height $h(t)$. The dashed green curve in Fig.~\ref{fg:parity} shows the inferred bouncing phase $\varphi(t) = \arctan(Q/I)$ of the droplet as a function of time. It reveals that the parity swap that may also be viewed as a `time switch' and controlled by the superlattice modulation is quantified by a total phase accumulation $\gamma_\tau=\pi$. The overshoots at the start and the restoration at the end of the modulation are repeatable physical effects caused by hysteresis when the bath acceleration is rapidly changed. The deterministic A-to-B parity swap can be induced on demand in two time directions by setting the sign of the rate of change of $\theta(t)$ in the drive, and this phenomenon permits the following interpretation. The A parity discrete time crystal melts when the long lattice ($|V_l|>0$) is mixed into the drive $a(t)$. The superlattice modulation of $a(t)$ then results in the droplet being pumped along the time lattice, and once the pumping stops ($V_l=0$) the droplet resolidifies into the B parity time crystal phase. 

\section*{Topological pumping}

\begin{figure*}[b!]
\includegraphics[width=1\columnwidth]{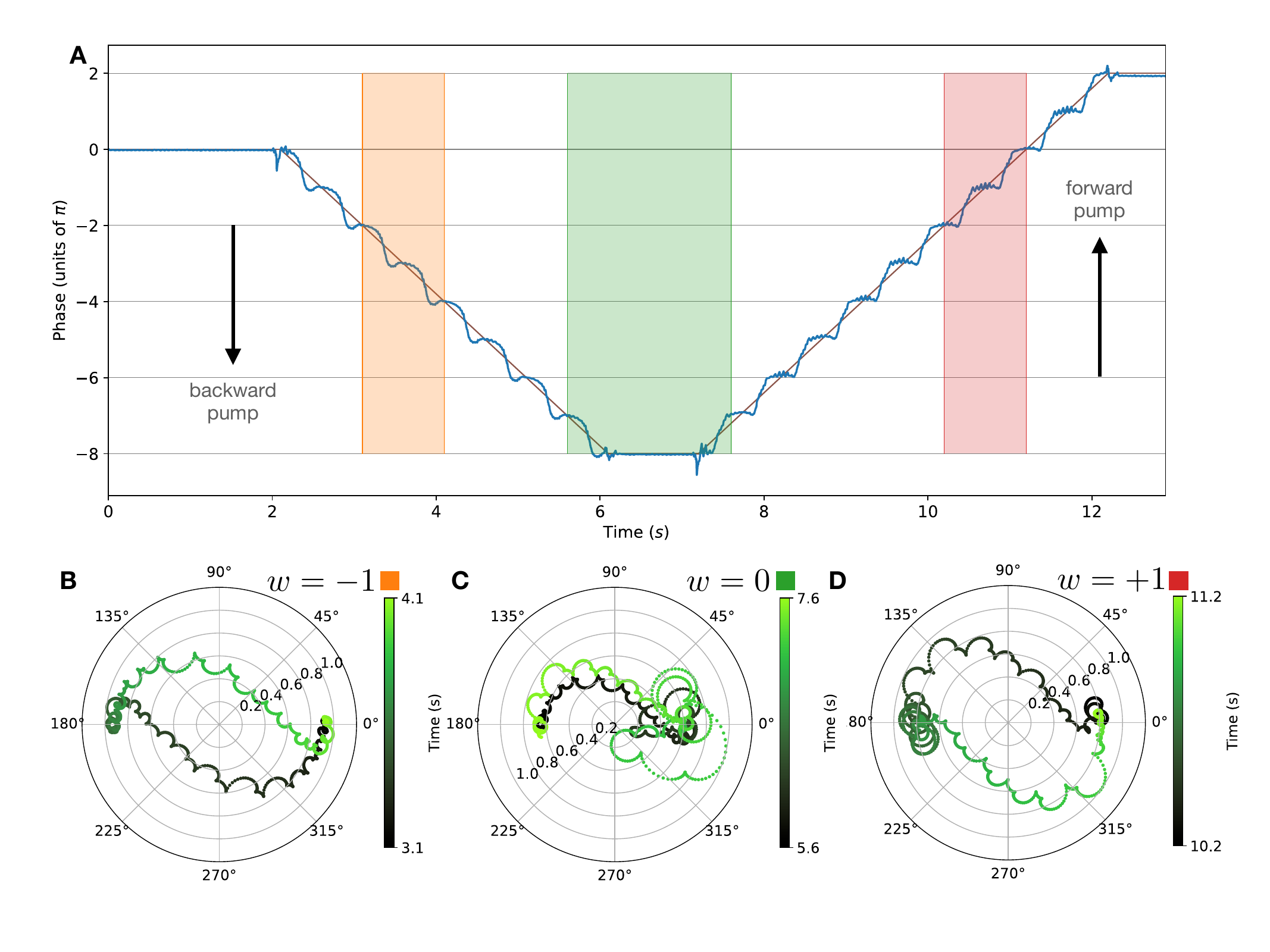}
\caption{{\bf Topological pumping.} (\textbf{A}), Phase evolution $\varphi(t)$ of the droplet for the case of 4 pump cycles backward in time followed by 5 cycles forward (blue) and the modulation function $\theta(t)$ of the drive (brown) as functions of time. (\textbf{B-D}) The skipping orbit trajectories $Z(t)$ for the time intervals indicated in ({A}) corresponding to topological indices -1, 0 and +1, respectively. The pump amplitudes are in all cases as in Fig.~\ref{fg:parity}.}
\label{fg:pump2D}
\end{figure*}

The robustness of the A-to-B conversion leads us to consider more general pumping protocols encoded via $\theta(t)$ to reveal the underlying topological origin of this discrete time crystal pump. Figure \ref{fg:pump2D}A shows the measured phase $\varphi(t)$ (blue) together with $\theta(t)$ (gray) as functions of time for the case where the pump operates in the reverse mode for four seconds and in the forward mode for five seconds. Evidently, after each one-second pump cycle the droplet accumulates a phase $\gamma_\tau=\pm 2\pi$ with the sign corresponding to the pumping direction in the time lattice. 

Our pumping scheme which drives the phase winding quantified in Fig.~\ref{fg:pump2D} is rooted in topology. Figure \ref{fg:pump2D}B-D uses a polar plane representation of the $Z(t)=(I(t),Q(t))$ phasor to quantify the accumulated time-geometric phase $\gamma_\tau$ for three particular time intervals specified by the shaded areas in Fig.~\ref{fg:pump2D}A. The topological index for the three cases (B-D) are inferred to be $w=-1,0,+1$, respectively. This winding number counts the number of times the trajectory $Z(t)$ encircles the origin with the sign corresponding to clockwise and anticlockwise windings. We emphasize that the accumulated geometric phase $\gamma_\tau$, and consequently the topological index $w$, do not depend on the speed of the pumping, or it being linear function of time, and is determined only by the geometry and topology of the trajectory of $\theta(t)$ through the parameter space. The paths in (B-D) can therefore be continuously deformed into arbitrary shapes by smoothly varying $a(t)$ and provided the path does not cross the origin, the winding number remains unaffected and is topologically protected. The winding number also precisely quantifies the integer number $2w$ of elementary time sites that the droplet is transported along the time lattice. In Fig.~\ref{fg:pump2D}A, the total pumped winding number and phase accumulation after nine complete pump cycles (four backwards and five forwards) are $w=1$ and $\gamma_\tau=2\pi$, resulting in quantised droplet transport of two time sites forward. We have verified the topological nature of the pumping by realising a broad range of pumps with varying speeds and total winding numbers (see Supplementary Figure~S2).

\section*{Synthetic time dimensions}

Figure \ref{fg:quanthall}A shows a two-dimensional (2D) representation (see Supplementary Figure~\ref{fg:1d2d}) of the pump process where the colour corresponds to the vertical position $h(t)$ of the droplet and the white trajectory is the curve in Fig.~\ref{fg:pump2D}A, facilitating multiple interpretations. First, the axis $T_1$ may be interpreted as a `position' coordinate of a 1D time lattice with $T_2$ corresponding to a `synthetic' time dimension. Within this interpretation at $T_2=0$ the green colour corresponds to the locations of the droplet(s) in a 1D time lattice at half-filling and the white trajectory corresponds to the `worldline' of a droplet as it is transported first backward and then forward along the time axis $T_1$ of the time lattice. Alternatively, this image represents a snapshot of the droplet density in a 2D time-time lattice spanned by time axes $T_1$ and $T_2$ with green and orange colour corresponding to occupied and empty sites, respectively. The pumping regions are penetrated by a `magnetic flux' $B_\perp$ (pink and green in Fig.~\ref{fg:quanthall}B and C) perpendicular to the 2D lattice plane causing a dramatic change in the droplet dynamics.

\section*{Relation to Foucault pendula and Laughlin pumps}

\begin{figure*}[b!]
\includegraphics[width=1\columnwidth]{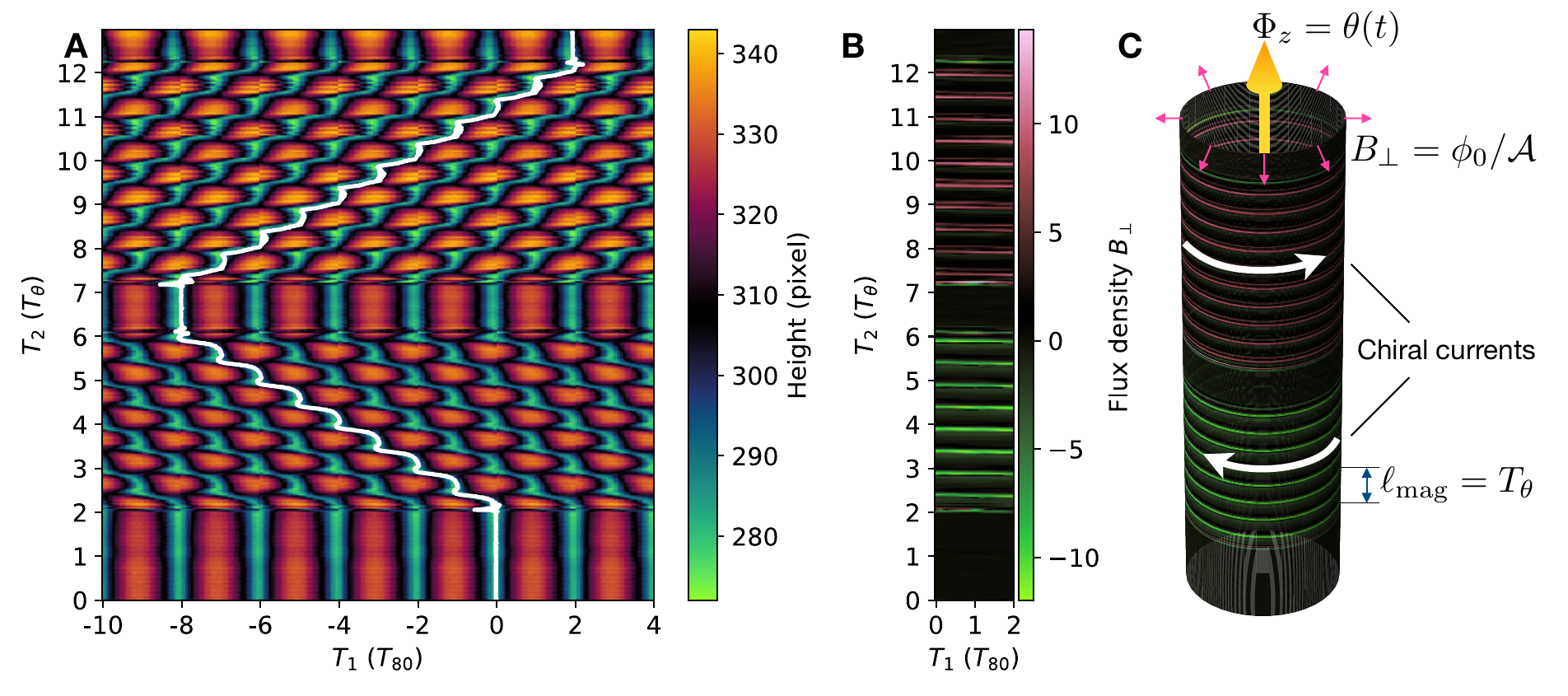}
\caption{{\bf Quantum Hall perspective.} (\textbf{A}), Droplet transport in an effective time lattice parametrised by $T_1$ and $T_2$ (see Supplementary Fig.~\ref{fg:1d2d}). The white trajectory shows $\varphi(t)$ as in Fig.~\ref{fg:pump2D}{A} with change of axes $T_1(T_{80})\to \varphi(t)$ and $T_2(T_\theta)\to {\rm Time}(s)$, and corresponds to the temporal position of the droplet in the time lattice. The two pumping regions are pierced by a perpendicular magnetic field of strength $B_\perp = \phi_0/\mathcal{A}$, where $\mathcal{A}=T_{80}T_\theta$ is the `magnetic area'. (\textbf{B}) Flux density penetrating the time lattice of (A). (\textbf{C}) Wrapping of (B) into a cylinder to illustrate the connection to quantum Hall phenomenology using Laughlin's thought experiment. The white and pink arrows respectively denote the chiral currents and the radial magnetic field. The yellow arrow represents fictious magnetic flux through the cylinder. Every time the axial flux increases by one flux quantum $\phi_0$, the droplet `charge' is pumped one magnetic length $\ell_{\rm mag}$ forward in the axial direction, wrapping around the cylinder once and crossing two flux rings.}
\label{fg:quanthall}
\end{figure*}

It is insightful to assign each sample of the measured time series $h(t)$ a normalised 2D phasor $(I(t),Q(t))/|(I(t),Q(t))|$ and reshape it into a 2D vector field ${\bf{A}}(T_1,T_2)$ in the same fashion as the 2D scalar field $h(T_1,T_2)$ of Fig.~\ref{fg:quanthall}A is reshaped from $h(t)$ (see Supplementary Fig.~S\ref{fg:1d2d}). Figure \ref{fg:quanthall}B shows the projected flux density $B_\perp = (\nabla\times{\bf{A}})\cdot {\bf z}$ where ${\bf z} = {\rm sign}({I(t)})\hat{\bf{e}}_z$, such that every crossing of a `flux line' (green or pink stripe) corresponds to a directional parity swap in the course of the discretised droplet transport. Similarly with the spatial quantum Thouless pumps that are characterised by a geometric Berry phase and a topological Chern number \cite{Citro2023a}, this perspective furnishes an analogy of the discrete time crystal pump with an integer quantum Hall-like effect in the synthetic 2D time-domain, characterised by $\gamma_\tau$ and $w$. Specifically, wrapping the Fig.~\ref{fg:quanthall}B into a Corbino disk or a cylinder by identifying the ends of the periodic $T_1$ axis, one arrives at the viewpoint of a time-domain `Laughlin pump' \cite{Laughlin1981a,Fabre2022a} illustrated in Fig.~\ref{fg:quanthall}C where the  droplet current has a chiral component in the azimuthal direction and an axial component $I_z$ that crosses rings of flux (green and pink). When the fictious Laughlin pump flux $\Phi_z=\theta(t)=\phi_0 t/T_\theta$ (yellow arrow) increases by one flux quantum $\phi_0=2\pi$, the droplet `charge' $q$ is pumped one `magnetic length' $\ell_{\rm mag}=T_\theta$ forward in the axial $T_2$ direction resulting in a current $I_z$. Using the quantum Hall terminology, the associated Hall conductivity is thus $\sigma_{\rm H} = I_z/\partial_t\Phi_z=\nu q/\phi_0$, where $\nu q=I_zT_\theta$ quantifies the integer number $\nu$ of charges transported in time $T_\theta$, and $q/\phi_0$ is the conductance quantum. With this quantum Hall perspective, following the prescription used for neutral atoms in rotating traps~\cite{Fletcher2021a,Mukherjee2022a}, the skipping orbit trajectories evident in Fig.~\ref{fg:pump2D}B-D may be broken down into a slow motion of the guiding centre $(X,Y)=(I-\xi,Q-\eta)$ coordinates and fast motion of the cyclotron $(\xi,\eta)=(I-X,Q-Y)$ coordinates with respective characteristic frequencies $\omega_g = 2\pi/T_{\theta}$ and $\omega_c= 2\pi/T_{80}$. This establishes the underlying anholonomy due to the non-commutative spaces $[X,Y]=-[\xi,\eta]=\mathcal{A}$, where $X=I/2+\mathcal{A}\partial_Q$, $Y=Q/2-\mathcal{A}\partial_I$, $\xi=I/2-\mathcal{A}\partial_Q$, and $\eta=Q/2+\mathcal{A}\partial_I$. The `magnetic area' $\mathcal{A}=T_{80}T_{\theta}$ defines the minimum area (uncertainty) spanned by the non-commutative phase space variables. The unusual inversion of the skipping orbits observed in Fig.~\ref{fg:pump2D}D where the tips point radially inward rather than outward as is the case in b originates from the sign change of $T_\theta$, which flips the sign of the commutators. In the Foucault pendulum the minimum phase space area is proportional to $L/g$ \cite{Somerville1972a},
where $L$ is the length of the pendulum, and in quantum systems it is proportional to the Planck constant $\hbar$. Similar skipping orbit inversion could therefore be observed in such systems if it was possible to change the sign of $L/g$ and $\hbar$, respectively.

\section*{Discussion}

In spatial quantum Thouless pumps the particle transport relies on quantum mechanical tunneling between the neighbouring spatial lattice sites. Because the minima of the lattice sites do not move during the pumping process,  a classical particle would remain trapped in its original site for all times \cite{Lohse2016a,Nakajima2016a,Lu2016a}. It is therefore remarkable that for our time-domain pumping scheme, in which the minima of the time lattice sites also do not `move', topological transport can be facilitated by completely classical means. To conclude, we have realised topological pumping of fluid droplets and have deployed such pumps for quantised transport of discrete classical time crystals in the time-domain. Adapting our pumping protocol to quantum systems may lead to novel applications such as topologically protected time keeping.



\bibliographystyle{Science}

\section*{Acknowledgments}
\noindent N.K. thanks Matthew Chilcott for technical assistance. 
\noindent{\bf Funding:} N.K. was supported by the Marsden Fund of New Zealand (Contract No. UOO1923). T.S. was supported by Australian Government through the Australian Research Council (ARC) Future Fellowship FT180100020. T.P. was supported by the European Research Council (ERC) under the European Union's Horizon 2020 research and innovation programme (Grant Agreement No.~101019739-LongRangeFermi).


\clearpage{}



\clearpage

\section*{Materials and Methods}
\noindent\textbf{Experimental setup:}
Main components of our experiment have been described in \cite{Simula2023a}. Added features include a tracking laser beam aimed at the droplet, allowing us to extract its vertical dynamics, and a digital arbitrary waveform generator (AWG) to drive the shaker. The driving waveforms $a(t)$ are synthesized numerically at a $100$ kHz sampling rate and uploaded to the memory of the AWG that feeds the signal to a linear amplifier that drives the motion of the electrodynamic shaker and the fluid bath mounted on top of the shaker. The topological pumps are embedded into a $17$ s long 80 Hz sinusoidal waveform with phase matched ends, and the full waveform is continuously looped. We continuously measured the RMS acceleration of the bath to adjust the overall voltage gain of the drive to set $V_s$ and $V_l$. A second channel of the AWG is used for triggering the high-speed camera aquisition sequence synchronously with the bath drive.

The droplets are produced using a piezoelectric droplet generator with a 0.6 mm nozzle diameter. The actual droplet diameter is slightly larger, typically 0.73 mm, and is inferred optically. 
The droplet is trapped in the horizontal plane by a circular well of 4.7 mm depth and 10.7 mm diameter partially filled with 20 cSt silicone oil. The oil forms a smooth meniscus with a positive curvature and a bouncing oil droplet will experience a planar restoring force toward the centre of the well where it remains horizontally trapped.

\section*{Supplementary Text}
\noindent\textbf{IQ demodulation:}
To extract the phase $\varphi(t)$ of the droplet, we perform an in-phase and quadrature demodulation of the bouncing signal $h(t)$ according to
\begin{eqnarray*}
I(t) =&\int_t^{t+2T}\cos(\pi f t') h(t')dt' /h_0  \\
Q(t) =&\int_t^{t+2T}\sin(\pi f t') h(t')dt' /h_0,
\end{eqnarray*}
where the normalisation constant $h_0$ is of the order of the droplet diameter and sets the scale for $I$ and $Q$ giving them the dimension of time.
The droplet phase is then calculated as $\varphi(t)=\arctan(Q/I)$. The absolute phase $\varphi(t)$ is not gauge-invariant quantity since the time origin may be chosen arbitrarily. By contrast, the integral 
\begin{eqnarray*}
\gamma_\tau=\int_{t_i}^{t_f} \frac{\partial \varphi(t')}{\partial t'} dt'
\end{eqnarray*}
over a time interval $\tau=t_f-t_i$ is gauge-invariant geometric phase and corresponds to the Berry--Hannay angle\cite{Berry1985a,Hannay1985a}. For time intervals $\tau$ that satisfy $\theta(t+\tau) = \theta(t)$, $\gamma_\tau=w2\pi$.


\noindent\textbf{Time crystal stability and the absence of time phonons:}
To demonstrate the extreme stability of our time crystals, we sampled a single time crystal over a $3\,000$s time interval corresponding to more than $120\,000$ period doubled oscillations, enabled by the 10MHz internal clock of the AWG that allows the camera trigger to remain synchronised to the drive for long periods of time. We collected 42 samples at 800 fps, each sample being 1s in duration. Supplementary Fig.~\ref{fg:stability}A shows the droplet bouncing as a function of time in a single 1s sample. The same color (blue, orange) markers are spaced at intervals of 1/40s, and the different colour markers are spaced at 1/80s intervals. Supplementary Fig.~\ref{fg:stability}B shows all 42 samples graphed on a single one second time axis revealing that the droplet bouncing remains phase locked to one and the same parity bouncing for all times as illustrated in Supplementary Fig.~\ref{fg:stability}C, which captures the same data as b but using the full extent of the time axis. We attribute this robustness to the absence of `time phonon' excitations, or more precisely them being overdamped. By considering a plain $a_{80}$ driving and sinusoidally modulating its phase, we are able to induce these time phonon phase oscillations onto the droplet as shown in Supplementary Fig.\ref{fg:stability}D (droplet height) and e (droplet phase). We have considered a range of phonon excitation frequencies and in all cases when the excitation is halted, the droplet phase also stops oscillating abruptly. Our interpretation of this is that the time phonons are overdamped and are not able to propagate in these time lattices and that this absence of phonon excitations underlies the remarkable stability of the droplet time crystal.

\noindent\textbf{Pump speed:}
To further highlight the topological origin of the pumping process, we have realised pumps that vary by factor of 160 in their speed. Supplementary Fig.~\ref{fg:speed} shows the accumulated phase as a function of time for 8 different pump speeds as functions of scaled time $T_\theta$. In each case, the drive phase $\theta(t)=2\pi t/T_\theta$. We have also used nonlinear functions of time for $\theta(t)$ to further confirm the topological robustness of the pumping process.

\noindent\textbf{Pump parameters:}
The drive parameters for Fig.~\ref{fg:parity} are: for $t<12s$, $V_s=2.93 g, V_l=0$; for $12.0s<t<12.05s$, $V_s=2.93g- 1.65g(t-12s)/0.05s,V_l=0.46g(t-12s)/0.05s$, $\theta(t)=0$; for $12.05s<t<13.05s$, $V_s=1.28g,V_l=0.46g$, $\theta(t)=2\pi (t-12.05s)$Hz; for $13.05s<t<13.1s$, $V_s=1.28g + 1.65g(t-13.05s)/0.05s,V_l=0.46g(1 - (t-13.05s)/0.05s)$, $\theta(t)=0$; and for $t>13.1s$, $V_s=2.93 g, V_l=0$.

\noindent\textbf{Tables:}
The filename of each .csv file within the compressed TableS1.zip file refers to the corresponding figure label. In each file the first dimension provides the values on the abscissa and the subsequent dimensions provide the values on the ordinate, except in the cases of Fig3.B-D where the two dimensions are the polar angle and radial coordinate.

\noindent\textbf{Movie:}
Video of the 40 Hz bouncing motion of the droplet time crystal corresponding to Fig.~\ref{fg:setup}. The red color is a reflection from the tracking laser and the white light is from a back illumination LED and reveals the waves produced by the droplet on the fluid surface. The 80 Hz bath motion is visible in the four corners of this video.


\clearpage

\setcounter{figure}{0} 
\renewcommand{\figurename}{{\bf Fig.~S\hspace{-4pt}}}
\begin{figure*}[]
\centering
\includegraphics[width=1\textwidth]{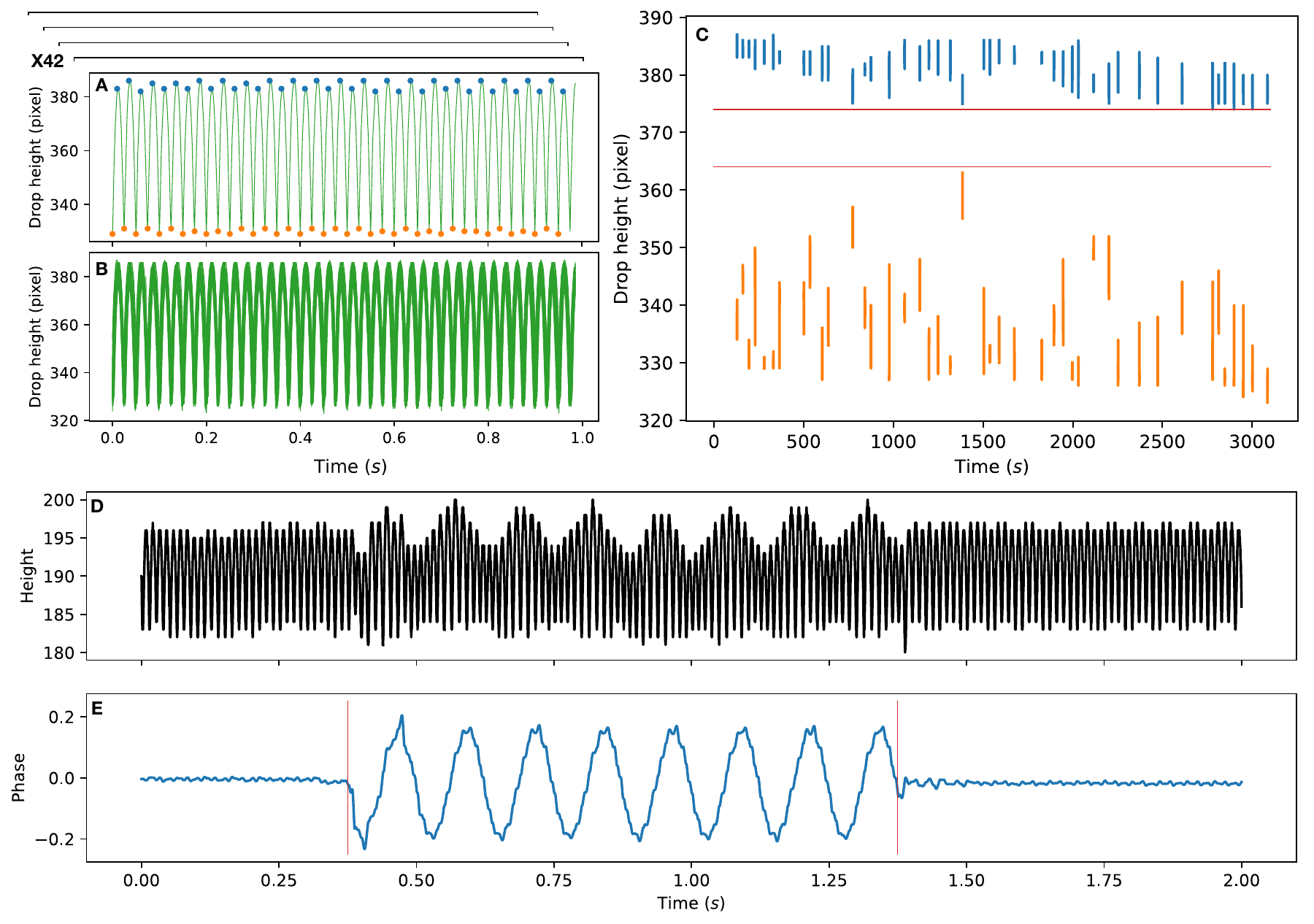}
\caption{{\bf Time crystal stability.} (\textbf{A}) One second snapshot showing the droplet height as a function of time (green continuous curve). The blue and orange markers show the same data sampled at 80 fps with an alternating colour. (\textbf{B}) 42 overlaid one-second snapshots sampled from a single droplet over a 3000s duration at times shown in (\textbf{C}) comprising of more than $120\,000$ time crystal oscillation periods. Each orange and blue `stripe' in (C) corresponds to one of the 42 randomly timed snapshots with the green continuous curve removed. The horizontal red lines are drawn to illustrate that the gap between the blue and orange data points never closes due to persistent phase coherence of the bouncing. (\textbf{D}) (droplet height) and (\textbf{E}) (droplet phase) show the result of driving a phonon excitation by sinusoidally modulating the phase of a pure $a_{80}$ drive. When the excitation is stopped the phonon phase oscillation also ceases (E) instead of continuing to propagate. The vertical lines mark the start and the end of the phonon excitation in the driving waveform.}
\label{fg:stability}
\end{figure*}

\begin{figure}[]
\centering
\includegraphics[width=0.5\textwidth]{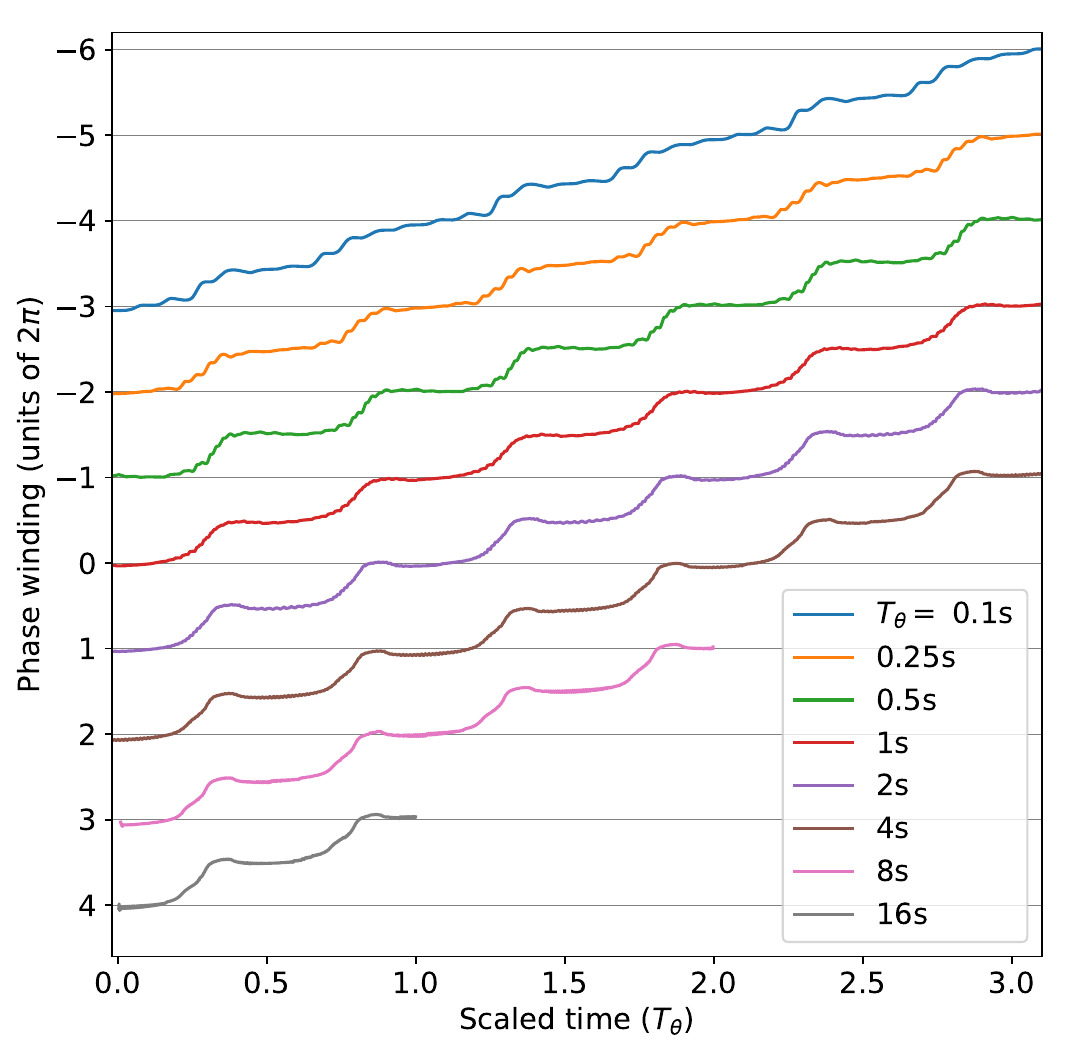}
\caption{{\bf Varying pump speed.} Accumulated winding number as functions of time scaled by $T_\theta$. For $T_\theta=1.0$s one pump cycle takes one second. The fastest $T_\theta=0.1$s and the slowest $T_\theta=16$s pumps differ by factor of 160 in their speed yet they capture qualitatively the same physics. This highlights that the pumping is topological and the accumulated winding does not depend on how rapid the pumping is. In the case $T_\theta=0.1$s the droplet only bounces a handful of times during one pump cycle so it is remarkable that the pumping still proceeds as if it was adiabatic. The pump amplitudes are in all cases as in Fig.~\ref{fg:parity}. }
\label{fg:speed}
\end{figure}

\begin{figure*}[]
\centering
\includegraphics[width=0.8\textwidth]{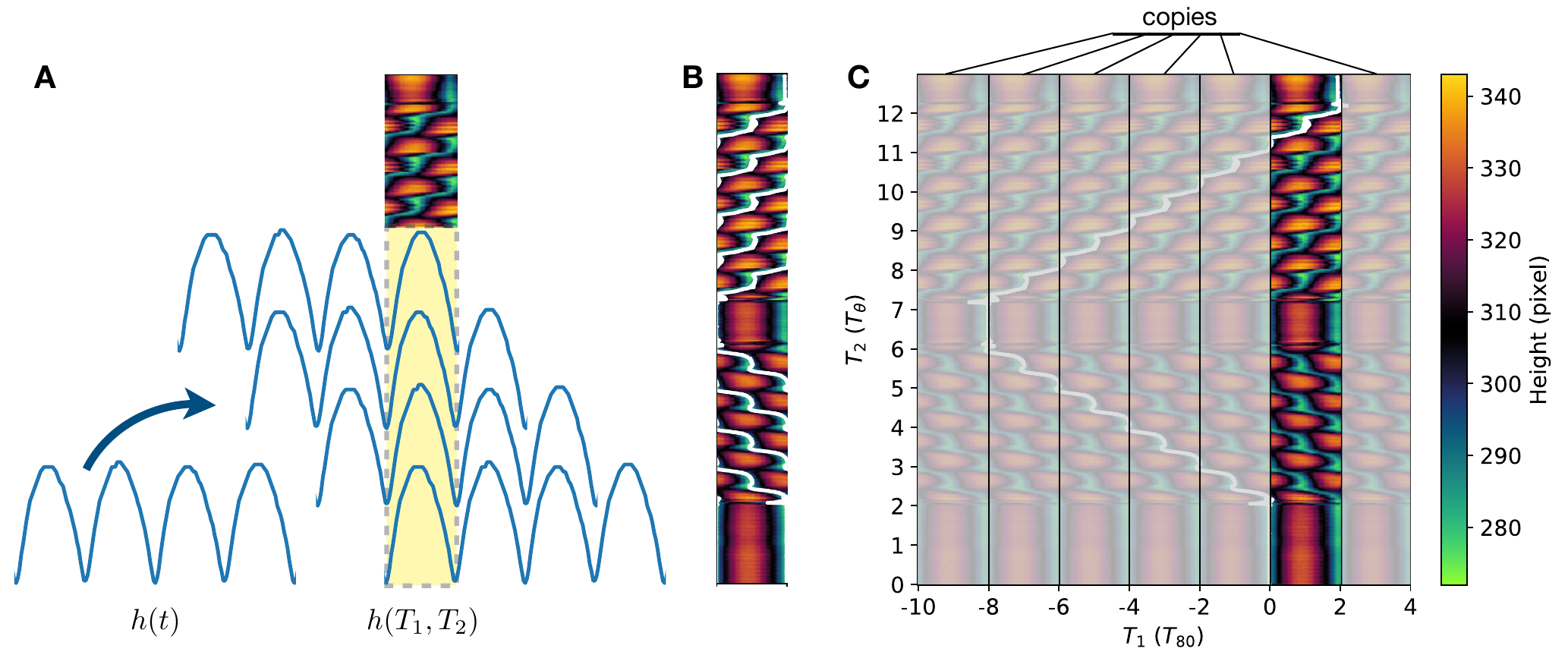}
\caption{{\bf Construction of Fig.~\ref{fg:pump2D}A.} The measured time series $h(t)$ is a 1D $N\times 1$ array, which is first sequenced into sections of $2T_{80}$ duration and then reshaped into a 2D matrix $h(T_1,T_2)$ with $N$ elements. (\textbf{B}) Plot of $h(T_1,T_2)$ with the measured phase $\varphi(t)$ shown (white `stripes'). (\textbf{C}) The 2D data $h(T_1,T_2)$ is phase unwrapped by padding a larger matrix with copies of $h(T_1,T_2)$, and finally the unwrapped phase function $\varphi(t)$ is overlaid.}
\label{fg:1d2d}
\end{figure*}




\end{document}